\documentclass[aps,pra,reprint,groupedaddress,amsmath,amssymb,preprintnumbers,showpacs]{revtex4-1}
\usepackage[normalem]{ulem}

\usepackage{enumerate}
\usepackage{graphicx}
\usepackage{color}
\usepackage{float} 
\usepackage{xspace}
\graphicspath{{./}}

\newcommand{\ket}[1]{\bigl| #1 \bigr>} 
\newcommand{\bra}[1]{\bigl< #1 \bigr|} 
\newcommand{\braket}[2]{\bigl< #1 \vphantom{#2} \bigr|
 \bigl. #2 \vphantom{#1} \bigr>} 
\newcommand{\abs}[1]{\left| #1 \right|} 

\newcommand{\mol}{$\text{H}_2^+$\xspace}

\begin{document}
\title{Inter- and Intracycle Interference Effects in Strong-Field Dissociative Ionization}
\author{Lun \surname{Yue}}
\author{Lars Bojer \surname{Madsen}}
\affiliation{Department of Physics and Astronomy, Aarhus University, DK-8000 Aarhus C, Denmark}
\date{\today} 

\begin{abstract}
We theoretically study  dissociative ionization of \mol exposed to strong linearly polarized few-cycle visible, near-infrared and near-midinfrared laser pulses. 
We find rich energy-sharing structures in the combined electron and nuclear kinetic energy spectra with features that are a priori at odds with simple energy conservation arguments.
We explain the structures as interferences between wave packets released during different optical cycles, and 
during the same optical cycle, respectively. Both inter- and intracycle interference structures are clearly visible in the joint energy spectra. 
The shapes of the  interference structures depend 
on the dynamics leading to the double continuum, and carry sub-femtosecond information.
\end{abstract}

\maketitle

\begin{figure} 
  \centering
  \includegraphics[width=0.5\textwidth]{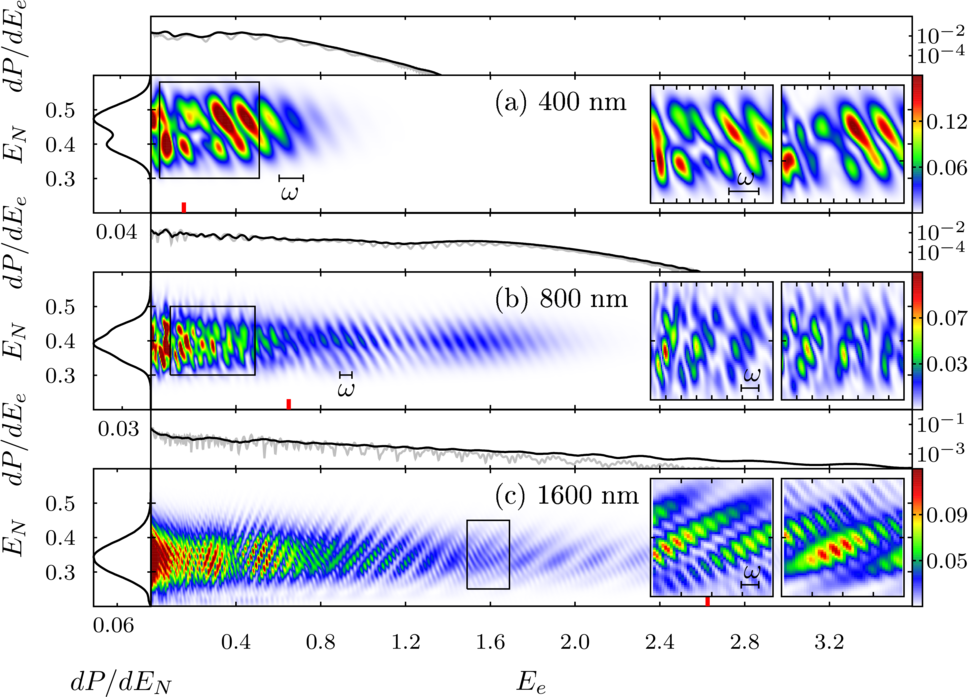} 
  \caption{(color online). JES  of \mol in units of a.u.$^{-2}$ after exposure to laser pulses with $I=3\times 10^{14}$ W/cm$^2$, $N_c=6$, and (a) $\lambda=400$ nm (b) $800$ nm (c) $1600$ nm, calculated by TDSE. The side and upper subpanels show nuclear spectra and PES in units of a.u.$^{-1}$. The grey lines shows  the fixed-nuclei PES with $R=2.06$ with rescaled magnitude. The left and right insets show zooms of the JES in the rectangular box, for electrons escaping in the positive and negative $x$-direction, respectively. In the insets, the color-bar range is from zero to 0.15 (top), 0.125 (middle), and 0.021 (bottom). The red markers indicate values of $U_p$. 
    Units for $E_e$ and $E_N$ are a.u.
}
  \label{fig:1}
\end{figure}

Strong-field ionization is of fundamental interest, as it constitutes the first step in a range of processes including high-harmonic generation and rescattering ionization \cite{Krausz09}.
A characteristic atomic strong-field photoelectron spectrum (PES) features peaks 
separated by the photon energy, and the corresponding process is referred to as above threshold ionization (ATI) \cite{Agostini79}. The appearance of such structures in the PES can be interpreted as interference between electronic wave packets (WPs) released during different cycles of the pulse, referred to as intercycle interferences. 
A decade ago, it was 
demonstrated  that the PES can exhibit structures with modulations on larger energy scales than that of the photon,
 which were interpreted as intracycle interferences between WPs released during the same subcycle \cite{Lindner05,Gopal09,Arbo06}. 
The signatures of intracycle interferences have been difficult to identify in the complex spectra following strong-field ionization of atoms by linearly polarized laser pulses, but
a two-color scheme with orthogonally polarized pulses very recently allowed a unique identification in  Ar~\cite{Richter15}. 
Subcycle interference effects in molecules were discussed in connection with strong-field electron holography \cite{Bian12,Meckel14}. In the theoretical treatment of such studies, the fixed-nuclei approximation was applied, and nuclear motion were neglected. In the case of dissociative ionization (DI) of small molecules, it was, however,  predicted \cite{Madsen12,Silva13} and  verified \cite{Wu13} that correlation 
 with the nuclei cannot be neglected, and nuclear motion therefore has to be included in the description. 
 In the joint energy spectra (JES) of the electron $E_e$ and nuclear $E_N$ energies,
 diagonal maxima separated by the laser frequency, $\omega$, appear in the continuum JESs showing the energy sharing between electron and nuclei, 
[atomic units (a.u.) used unless stated otherwise] 
 \begin{equation}
 \label{energy}
 E_e + E_N = E_0 + n\omega - U_p,
 \end{equation}
 where $E_0$ is the bound state energy, $U_p$ the ponderomotive energy, and $n$ the number of absorbed photons [Figs. \ref{fig:1}(a) and \ref{fig:1}(b)].
The traditional ATI peaks in the PES were shown to be either badly resolved or gone completely due to the energy sharing  \cite{Madsen12,Silva13,Wang16}.  For field parameters close to the tunneling regime, the electrons were claimed to not share the energy with the nuclei at all due to tunneling electrons \cite{Silva13}, leading to a distribution of $E_N$ for fixed $E_e$, i.e., for $E_e \simeq \text{const}$ [Figs.~\ref{fig:4}(a) and \ref{fig:4}(b)]. In this work, we show that the exact inclusion of the nuclear degree of freedom in ionization  by strong few-cycle laser pulses 
leads to clear interference structures in the JES that at first sight defy standard energy conservation interpretations
with pronounced cross-diagonal structures, i.e.,  maxima  approximately at $E_e- E_N \simeq \text{const}'$ [Fig.~1(c)]. 
In the JES with $E_N$ vs $E_e$ we then have (i) diagonal maxima with positive slope [Eq.~\eqref{energy}], 
(ii) vertical maxima at constant $E_e$ (infinite slope), and (iii) cross-diagonal maxima with negative slope.
We  show that all   
the structures (i)-(iii) are due to intra- and intercycle interference effects and the laser dependent DI dynamics.
 The main characteristics of the JES may be controlled by changing the frequency of the driving pulse. 
The interplay between intra- and intercycle structures is present in the JES for all wavelengths considered, 
but is especially clear using wavelength close to the midinfrared. 
Intense, midinfrared pulses are currently developed~\cite{Popmintchev12,Silva15,Wolter15,Duval15} stimulated by the $\lambda^2$ scaling of the HHG and ATI cutoffs of importance for  sub-attosecond pulse generation \cite{Tate07,HernandezGarcia13}, strong-field holography \cite{Huismans11,Bian11} and laser-induced electron diffraction \cite{Blaga12,Wolter15}.

We base our study on \mol, an archetypal molecule that has predicted a range of general strong field phenomena, including charge resonance enhanced ionization \cite{Zuo95}, above-threshold Coulomb explosion \cite{Esry06}, bond-softening \cite{Bucksbaum90}, -hardening \cite{Bandrauk81,Zavriyev93}, and enhanced dissociation \cite{Yue15}.
We consider a co-linear model that includes the electronic and nuclear dimension aligned with the linearly polarized pulse, with the Hamiltonian 
\begin{equation}
  \label{eq:1}
  H(t)=T_e+T_N+V_{eN}+V_N+V_I(t),
\end{equation}
with $T_e=-(1/2\mu)\partial^2/\partial x^2$, $T_N=-(1/m_p)\partial^2/\partial R^2$, $V_{eN}=-1/\sqrt{{(x-R/2)^2+a(R)}}-1/\sqrt{(x+R/2)^2+a(R)}$, $V_N=1/R$,  $V_I(t)=-i \beta A(t) \partial/\partial x $,  $m_p$ the proton mass, $\mu=2m_p/(2m_p+1)$, $\beta = (m_p+1)/m_p$, $x$ the electronic coordinate in relation to the center of mass of the nuclei, $R$ the internuclear distance, and $a(R)$ chosen to produce the exact $1s\sigma_g$ Born-Oppenheimer (BO) curve. The initial ground state $|\Psi_0 \rangle$ has energy $E_0=-0.5973$, dissociation limit $E_d=-0.5$, and equilibrium internuclear distance $R_0=2.06$.
The vector potential is chosen as 
$A(t)=\left(F_0/\omega\right) g(t)\cos[\omega (t-\tau/2)]$, with envelope $g(t)=\sin^2(\pi t/\tau)$, pulse duration $\tau=N_c2\pi/\omega$,  number of cycles $N_c$, and peak intensity $I=F_0^2$. 
 The dipole approximation holds for the laser pulses considered here \cite{Reiss08,Wolter15}.
We solve the time-dependent Schr\"{o}dinger equation (TDSE) numerically \cite{Feit82} with complex absorbers 
to remove the outgoing flux. The JES is extracted using the time-dependent surface flux method \cite{Tao12,Yue13}, which allows  a moderate simulation volume $\abs{x}\le 200$ and $R\le 80$ even for near-midinfrared pulses.

Figure~\ref{fig:1} shows the JES for \mol after exposure to fields from the visible to the near-midinfrared regime. For the 400 nm pulse in Fig.~\ref{fig:1}(a),  the characteristic diagonal peaks satisfy Eq.~\eqref{energy} \cite{Madsen12,Silva13,Wu13,Yue13}. Due to the short pulse, electrons detected at $\pm x$ have different JESs [Fig.~\ref{fig:1}(a) insets]. Modulations along each diagonal is also observed. In the 800 nm case [Fig.~\ref{fig:1}(b)], in addition to the diagonal peaks, oscillatory structures on larger energy-scales are discernible, e.g., a structure with maximum at $E_e=1.6$ and minima at $E_e=1.2$ and $E_e=2.0$. Cross-diagonal structures with a positive slope are also shown in the insets. 
 For the 1600 nm case in Fig.~\ref{fig:1}(c), the JES is, at first glance, completely dominated by the cross-diagonal structures, seemingly satisfying the nonintuitive energy-conservations $E_e-E_N=\text{const}'$.
The insets show diagonal peaks, now superimposed on the cross-diagonal structures. 
Although one can distinguish fast oscillations superimposed on slower ones in the fixed nuclei PES in Fig.~\ref{fig:1}(c), they are completely washed out in the PES for moving nuclei, proving that for real molecules, the traditional PES is a poor observable. 
As we will show in the following, the different structures in the JES are due to the interplay between subcycle and intercycle interferences.

In the strong-field approximation (SFA) \cite{Keldysh64,*Faisal73,*Reiss80,*Lewenstein94}, the transition amplitude for direct ionization to the double continuum reads (we set $\mu\equiv \beta\equiv 1$ for simplicity)
 $T_{p,k}=-i \int_0^\tau dt \bra{\phi_{p}(t)\chi_{k}(t)}\tilde{V}_I(t)\ket{\Psi_0(t)}$,
with $\tilde{V}_I(t)=-i  A(t) \partial/\partial x+A(t)^2/2$, $\ket{\Psi_0(t)}=\ket{\Psi_0}e^{-iE_0 t}$, $\ket{\phi_{p}(t)}=\ket{\phi_p}e^{-i\int^t dt'[p+A(t')]^2/2}$ the Volkov wave with $E_e=p^2/2$, and $\ket{\chi_{k}(t)}=\ket{\chi_{k}}e^{-ik^2t/m_p}$ the Coulomb wave solution to $T_N+V_N$ with $E_N=k^2/m_p$. We calculate $T_{p,k}$ in the saddle-point approximation,
\begin{equation}
  \label{eq:3}
  T_{p,k}=\sum_s M_{p,k}(t_s)e^{iS_{p,k}(t_s)},
\end{equation}
with $S_{p,k}(t)=\int^{t}\left\{[p+A(t')]^2/2+E_N-E_0\right\}dt'$, $M_{p,k}(t_s)=-i\braket{\phi_p\chi_k}{\Psi_0} \sqrt{2\pi i/ \ddot{S}_{p,k}(t_s)}\left[pA(t_s)+A^2(t_s)/2\right]$, and the complex times $t_s$ solutions of
\begin{equation}
  \label{eq:4}
  \dot{S}_{p,k}(t)=\frac{\left[p+A(t)\right]^2}{2}+E_N-E_0=0
\end{equation}
with $0<\operatorname{Re}(t_s)<\tau$ and $\operatorname{Im}(t_s)>0$, and $\dot{S}_{p,k}$ ($\ddot{S}_{p,k}$) denoting derivative (double-derivative) of $S_{p,k}$ w.r.t. time, $t$.
For an $N_c$-cycle $\sin^2$ pulse there are 
$2(N_c+1)$ solutions. The terms in Eq.~\eqref{eq:3} correspond to quantum paths for electrons and nuclei reaching the 
same final momentum pair $(p,k)$ but released at different times $\operatorname{Re}(t_s)$, leading to interference in the continuum. 
\begin{figure} 
  \centering
  \includegraphics[width=0.5\textwidth]{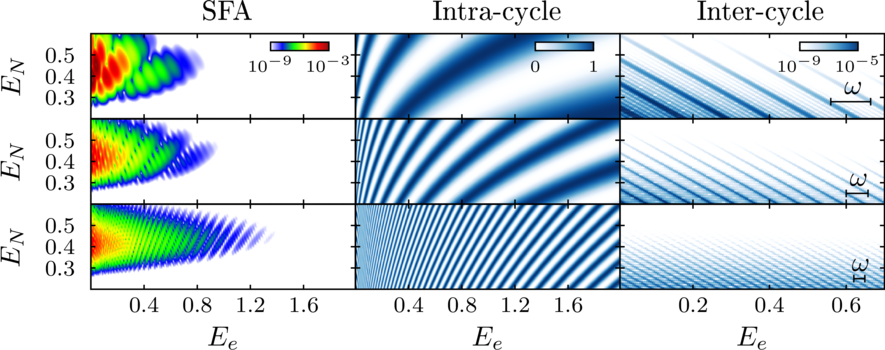} 
  \caption{(color online). Left panels: JES in a.u.$^{-2}$ calculated with SFA. Middle and right panels: Intracycle $G^\text{intra}_{p,k}$ and intercycle $G^\text{inter}_{p,k}$ contributions [Eq.~\eqref{eq:7}] assuming a constant field envelope. The pulse parameters are $I=3\times 10^{14}$ W/cm$^2$, $N_c=6$, and from top to bottom: $\lambda=$400 nm, 800 nm, and 1600 nm. All axes are in a.u.}
  \label{fig:2}
\end{figure}
The left panels of Fig.~\ref{fig:2} show the SFA JESs
obtained from $|T_{p,k}|^2$ \cite{Yue13,Yue14}.
We use a logarithmic scale for the SFA analysis due to its qualitative nature.
The qualitative similarities with Fig.~\ref{fig:1} are evident. For 400 nm, the SFA shows three lobes eminating in the cross-diagonal direction, with diagonal structures superimposed on each lobe, which are consistent with the modulations along the diagonals in Fig.~\ref{fig:1}(a). For 800 nm and 1600 nm in Fig.~\ref{fig:2}, the same trend is observed, with shorter spacings between the cross-diagonal structures as the wavelength is increased. 

\begin{figure} 
  \centering
  \includegraphics[width=0.5\textwidth]{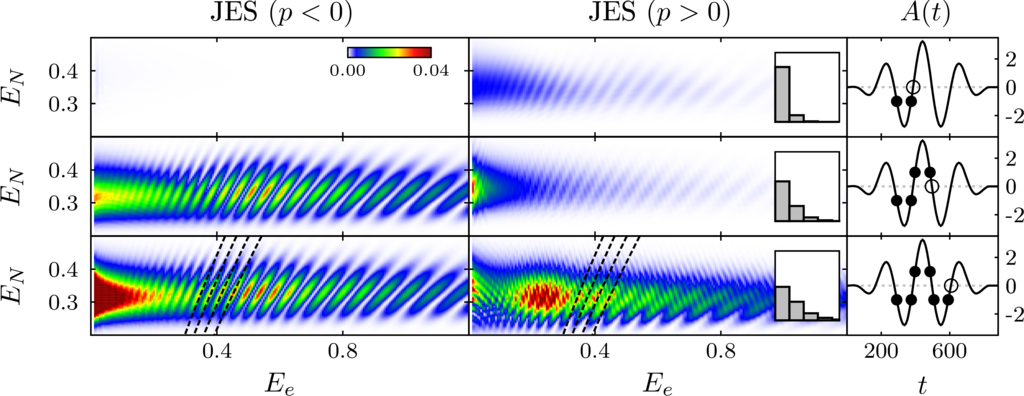} 
  \caption{(color online). The buildup of the JES for $\lambda = 1600$ nm, $I=3\times 10^{14}$ W/cm$^2$, and $N_c=4$, calculated with TDSE. The right panels depict the pulse, with the circles indicating the instantaneous times and the dots indicating the contributing $\text{Re}(t_s)$ corresponding to $E_e=0.5$. The left (middle) panels depicts the JES for electrons with $p<0$ ($p>0$).  The insets show the vibrational populations of the first four vibrational states with the ordinate scale $[0,1]$. The dashed lines in the lowest panels show the analytical predictions $\Delta S_{p,k}=2\pi n$ for $n=-35$ to $-32$. The JESs are in a.u.$^{-2}$, while all axes are in a.u. }
  \label{fig:3}
\end{figure}

\begin{figure} 
  \centering
  \includegraphics[width=0.5\textwidth]{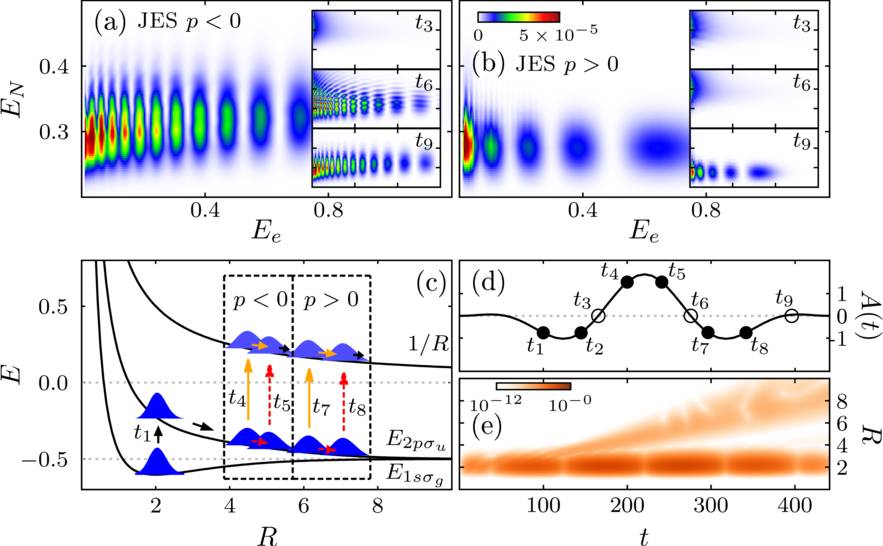} 
  \caption{(color online). (a) JES from TDSE for $\lambda=1600$ nm, $I=10^{14}$ W/cm$^2$, and $N_c$=2, with $p<0$. The insets show the time-resolved buildup of the JES during the pulse shown in (d). (b) Same as (a), with $p>0$. (c) Schematic of the nuclear dynamics. The three curves are the lowest two BO curves corresponding to the electronic states $1s\sigma_g$ and $2p\sigma_u$, and the DI curve $1/R$. (e) Nuclear density in the $2p\sigma_u$ electronic state using a two-surface model (see text). The JESs are in a.u.$^{-2}$, while all axes are in a.u. }
  \label{fig:4}
\end{figure}

For further analysis, we consider an $N_c$-cycle flat-top pulse with $A(t)=(F_0/\omega)\cos(\omega t)$ during the flat-top segment $t\in[0,\tau]$ to obtain analytical expressions and a simple physical picture. For $p>0$, the two complex solutions to Eq.~\eqref{eq:4} in the $j$'th cycle satisfying $\operatorname{Im}(t_s)>0$ are
\begin{equation}
  \label{eq:interintra6}
  \begin{aligned}
    t_{j1}&=\frac{1}{\omega}\left[ \cos^{-1}\left(-\kappa-i\gamma\right)+2\pi(j-1)\right],\\
    t_{j2}&=\frac{1}{\omega}\left[2\pi- \cos^{-1}\left(-\kappa+i\gamma\right)+2\pi(j-1) \right],
  \end{aligned}
\end{equation}
where we have defined the scaled momentum $\kappa=\omega p/F_0$ and the $E_N$-dependent Keldysh parameter $\gamma=\omega\sqrt{2(-E_0+E_N)}/F_0$. 
 The transition probability then factorizes into $\abs{T_{p,k}}^2 \approx 4 \abs{M_{p,k}}^2 G^\text{inter}_{p,k}G^\text{intra}_{p,k}$, 
\begin{equation}
  \label{eq:7}
  \begin{aligned}
    G^\text{inter}_{p,k}&=\abs{\sum_j^{N_c}e^{i\bar{S}_{p,k,j}}}^2=\abs{B_{p,k}}^2\left[\frac{\sin(N_cQ_{p,k}/2)}{\sin(Q_{p,k}/2)} \right]^2,\\
    \quad G^\text{intra}_{p,k}&=\cos^2\left(\frac{\Delta S_{p,k}}{2}\right),
  \end{aligned}
\end{equation}
with $B_{p,k}$ independent of $N_c$, $\bar{S}_{p,k,j}=\left[S_{p,k}(t_{j1})+S_{p,k}(t_{j2})\right]/2$, $Q_{p,k}=2\pi(-E_0+E_e+E_N+U_p)/\omega$, and $\Delta S_{p,k}=S_{p,k}(t_{j1})-S_{p,k}(t_{j2})$ independent of $j$. 
In Refs.~\cite{Arbo06,Arbo10}, such a factorization approach was employed for atomic systems where $\text{Im}(t_s)$ and $E_0$ were neglected. In our case, however, due to the dependence on $E_N$, $\text{Im}(t_s)$ must be taken into account. 
The explicit expressions for $\abs{B_{p,k}}^2$ and $\Delta S_{p,k}$ are
\begin{widetext}
  \begin{align}
      \abs{B_{p,k}}^2
      &=\left[\frac{1}{2}\left(\sqrt{2}\sqrt{\abs{z_+z_-}+\kappa^2+\gamma^2-1} +\abs{z_+} +\abs{z_-}\right) \right]^{-\frac{F_0^2}{\omega^3}(\kappa^2+1/2+\gamma^2)}
      \exp\left[\frac{\sqrt{2}}{4} \frac{F_0^2}{\omega^3}
        \left( \gamma A + 3\kappa C  \right)  \right], \label{eq:interintra10a} \\
      \Delta S_{p,k}
      &=\frac{F_0^2}{\omega^3}
      \left[
        -\left(\kappa^2+\frac{1}{2}+\gamma^2\right)\cos^{-1}\left(\frac{\abs{z_+}-\abs{z_-}}{2}\right) 
        +\frac{\sqrt{2}}{4}(3 \kappa A- \gamma C) \label{eq:interintra10b}
      \right],
    \end{align}
\end{widetext}
with $z_+=\kappa+1+i\gamma$, $z_-=\kappa-1+i\gamma$, $A=\sqrt{\abs{z_+z_-}-\operatorname{Re}(z_+z_-)}$, and $C=\sqrt{\abs{z_+z_-}+\operatorname{Re}(z_+z_-)}$.
The factor $G^\text{inter}_{p,k}$ can be interpreted as intercycle interference where wave packets (WPs) released during different cycles interfere, giving rise to multiphoton peaks as described by Eq.~\eqref{energy}. This is shown in the right panels of Fig.~\ref{fig:2}, where $G^\text{inter}_{p,k}$ is plotted for different wavelengths. The factor $G^\text{intra}_{p,k}$  arises due to the intracycle interferences between WPs released during the same subcycle, and is depicted in the middle panels of Fig.~\ref{fig:2}. The period of the intracycle structures increases with  $E_e$ for fixed $E_N$, and decrease with $E_N$ for fixed $E_e$. With increasing $\omega$, the energy period increases as well. The increase of period with $E_e$ and $\omega$ can be understood intuitively: larger $E_e$ and $\omega$ imply shorter time interval $\Delta t$ between WP-releases in a subcycle due to $p\approx -A\left[\text{Re}(t_s)\right]$, which in turn corresponds to a larger energy period $\Delta E_e\sim 1/\Delta t$.
Writing the preexponential factor in Eq.~\eqref{eq:interintra10a} as $x=\exp(\ln x)$ and taking the limit $\omega \rightarrow 0$ of the total resulting exponent, we obtain to second order in $\omega$ 
\begin{equation}
  \label{eq:8}
  \abs{B_{p,k}}^2=\exp\left({-\frac{2s^3}{3F_0}}\right) \exp\left[{\left(\frac{s^2}{5}-p^2\right)\frac{s^3\omega^2}{3F_0^3}}\right],
\end{equation}
with $s=\sqrt{2(-E_0+E_N)}$. The first factor is recognized as the exponential factor in the tunneling ionization rate~\cite{Landau91} and predicts the decay of yield with increasing $E_N$, while the second factor predicts decay of yield with increasing $E_e$. Both are consistent with the right panels of Fig.~\ref{fig:2}. For the intracycle structures, Eq.~\eqref{eq:interintra10b} gives to lowest orders in $\omega$ 
\begin{equation}
  \label{eq:9}
  \begin{aligned}
    \Delta S_{p,k}=&-\frac{\pi}{4}\frac{F_0^2}{\omega^3}+2p\frac{F_0}{\omega^2}-\frac{\pi(p^2+s^2)}{2}\frac{1}{\omega}\\
    &+\left(\frac{p^3}{3}+\frac{3ps^2}{2}\right)\frac{1}{F_0}-\frac{p^2s^2}{2}\frac{\omega}{F_0^2}.
    \end{aligned} 
\end{equation}
Setting $\Delta S_{p,k}=2\pi n$, we readily obtain an analytical expression for the maxima of the intracycle structures [see Fig.~\ref{fig:3}]. Note that the oscillating parts of $G^\text{inter}_{p,k}$ and $G^\text{intra}_{p,k}$ have increasing periods in the $\omega \rightarrow 0$ limit. For very small $\omega$ this leads to a washing out of any modulation in the spectra in accordance with the expectation from tunneling theory in the dc limit.

The interplay between inter- and intracycle interferences is directly confirmed by the time-resolved formation of the JES by TDSE calculations, shown in Fig.~\ref{fig:3}. During the first $7/4$ cycles of the pulse shown in the top panels of Fig.~\ref{fig:3}, appreciable ionization can only occur during the half-cycle indicated by the dots due to the scaling of $F$ in Eq.~\eqref{eq:8}. 
The ionized WPs can be considered a double-slit in time, giving rise to intracycle interferences in the direction $p>0$. 
Half a cycle later, we observe intracycle structures in Fig.~\ref{fig:3} with $p<0$.
 Near the end of the pulse, intercycle structures separated by $\omega$ are superimposed on intracycle structures for $p>0$ due to interference of WPs ionized in two different cycles. The clear increase in the JES yield is due to the significant population of excited vibrational states with a smaller ionization potential (IP), shown in the insets of Fig.~\ref{fig:3}. A good agreement is observed between the TDSE intracycle peaks and the prediction $\Delta S_{p,k}=2\pi n$, with $\Delta S_{p,k}$ in Eq.~\eqref{eq:9} (dashed lines in Fig.~\ref{fig:3}). 

So far, we have associated the diagonal structures to intercycle effects and the cross-diagonal structures to intracycle effects. At lower intensities however, a different picture emerges, which is depicted in Figs.~\ref{fig:4}(a) and \ref{fig:4}(b). Instead of intracycle structures with positive slope, the structures are now vertical, and the JES for $p>0$ is shifted towards smaller $E_N$ compared to the JES for $p<0$. The physical explaination is sketched in Fig.~\ref{fig:4}(c). At $t_1$, the instantaneous intensity is too low to allow significant direct ionization.
Instead, a dissociative WP is created on the $2p\sigma_u$ curve. This is confirmed by a separate calculation for the 
nuclear dynamics wherein we neglect the electronic continuum, and only take into account the lowest two BO surfaces~\cite{Giusti-Suzor95}. 
The population density in $2p\sigma_u$ is shown in Fig.~\ref{fig:4}(e) and a dissociative nuclear WP is seen created at 
$t_1$. At $t\approx t_2$, the dissociative WP has moved towards larger $R$, but the IP and $F$ are still unsuitable for 
appreciable ionization, resulting in the JESs at $t_3$ shown in the first inset of Figs.~\ref{fig:4}(a) and \ref{fig:4}(b). 
At $t\approx t_4$ and $t_5$, two electronic continuum WPs are created, giving rise to intracycle interferences for $p<0$. 
However, as the $2p\sigma_u$ and $1/R$ curves are nearly parallel, the effective IP is almost independent of $R$, and 
$\Delta S_{p,k}$ is independent of $E_N$. This explains the vertical structures observed in Fig.~\ref{fig:4}(a). 
The nuclear WPs follow the two pathways indicated in Fig.~\ref{fig:4}(c) by the orange and red arrows, and lead to the 
same final nuclear WP. 
The average nuclear energy $\bar{E}_N$ of the final DI WP is the sum of $1/R(t_5)$ and the kinetic energy gained during dissociation from $t_1$ to $t_5$. The dissociative WP is created at $R\approx 3$ [Fig.~\ref{fig:4}(e)], which leads to $\bar{E}_N=0.289$, in good agreement with Fig.~\ref{fig:4}(a). 
By the same reasoning, the electronic WPs created at $t=t_7$ and $t=t_8$ interfere for $p>0$, resulting in the vertical intracycle interference patterns in Fig.~\ref{fig:4}(b) and a nuclear energy $\bar{E}_N=0.282$. 
For longer pulses, we observed near-vertical structures separated by $\omega$, corresponding to intercycle interferences (not shown).
Note that our understanding of the vertical structures are different from the one given in Ref.~\cite{Silva13}, where it was claimed that the vertical structures in the JES were due to tunneling electrons from the ground electronic state $1s\sigma_g$. The importance of the excited electronic states in the JES were discussed recently in Ref.~\cite{Mosert15}.

In conclusion, we solved the TDSE for a H$_2^+$ model without the BO approximation, and established that the rich structures observed in the JES, involving (i) diagonal [Eq.~\eqref{energy}], (ii) vertical, and (iii) cross-diagonal patterns, are due to the interplay between inter- and intracycle interferences.  
We characterized the $\lambda$, $N_c$, and $I$-dependence of these structures and used an SFA model including nuclear kinetic energy for their interpretation.
As strong-field physics in the long wavelength regime is
the physics
of valence electrons, our results should hold for multielectron diatomics as well.
 Indeed, note that the factors determining the interference patterns, $G_{p,k}^\text{intra}$ and $G_{p,k}^\text{inter}$ in Eq.~\eqref{eq:7}, depend on the specific molecular system only through $E_0$. 
 The PES alone provides a poor observable for the detection of the inter- and intracycle structures. Luckily, measurement techniques such as the cold-target recoil ion momentum spectroscopy \cite{Ullrich03,Wolter15} exist that can measure ions and electrons in coincidence, and  midinfrared sources are also already available \cite{Wolter15}.
The linear scale of the colorplots, and the distinct characteristic crossings of the structures (i)-(iii), therefore points to the feasibility of experimental verifications of the predicted effects. 
We note that the nuclear motion occurs on the femtosecond time-scale, while the intra-cycle effects happen on few-fs or sub-fs time-scales. Thus, the clear signatures of the intracycle effect in the JES reflects the strong electron-nuclear correlation. 
The inter- and intracycle interference structures  carry direct information of femtosecond and sub-femtosecond dynamics, and the present insights therefore  add to the understanding of time-dependent phenomena.

This work was supported by the Danish Center for Scientific Computing, an ERC-StG (Project No. 277767 - TDMET), and the VKR center of excellence, QUSCOPE.

 
%

\end{document}